\documentclass[aps,prd,groupedaddress,showpacs,notitlepage]{revtex4-1}
\usepackage{amsmath,amstext,amsbsy,amssymb}
\usepackage{bm}
\usepackage{color}

\newcommand{\ssA}{{\scriptscriptstyle{ A}}}
\newcommand{\ssV}{{\scriptscriptstyle{ V}}}
\newcommand{\ssD}{{\scriptscriptstyle{ D}}}

\newcommand{\ssR}{{\scriptscriptstyle{ R}}}

\newcommand{\ssB}{{\scriptscriptstyle{ B}}}
\newcommand{\ssL}{{\scriptscriptstyle{ L}}}

\newcommand{\ssh}{{\scriptscriptstyle{1/2}}}

\newcommand{\sschi}{{\scriptscriptstyle{ \chi}}}
\newcommand{\ssFD}{{\scriptscriptstyle{ FD}}}
\newcommand{\sse}{{\scriptscriptstyle{e}}}

\newcommand{\sspp}{{\scriptscriptstyle{1}}}
\newcommand{\ssmm}{{\scriptscriptstyle{-1}}}
\newcommand{\fV}{f_\ssV}
\newcommand{\fA}{f_\ssA}

\long\def\symbolfootnote[#1]#2{\begingroup%
	\def\thefootnote{\fnsymbol{footnote}}\footnote[#1]{#2}\endgroup}

\begin{document}
	
	\title{Semiclassical transport equations of Dirac particles in rotating frames }

	\author{\"{O}mer F. Dayi}
	%\email{dayi@itu.edu.tr }
	\author{Eda Kilin\c{c}arslan}
	%\email{kilincarslan@itu.edu.tr }
	\affiliation{%
		Physics Engineering Department, Faculty of Science and
		Letters, Istanbul Technical University,
		TR-34469, Maslak--Istanbul, Turkey}
	
\begin{abstract}
We establish  covariant semiclassical transport equations of massive spin-1/2 particles which are  generated by the  quantum kinetic equation modified by enthalpy current dependent terms. The purpose of  modification is to  take into account the noninertial properties due to the angular velocity of rotating frame which is equivalent to the fluid vorticity in the hydrodynamical  approach. We  present the equations satisfied by the Wigner function components and by studying their solutions in the semiclassical approximation we accomplish the transport equations.To acquire  a  three-dimensional kinetic theory, the relativistic kinetic equations in the comoving frame are integrated  over the zeroth component of four-momentum.   The  resulting  vector and axial-vector currents are calculated at zero temperature. There exists another  three-dimensional formulation of Dirac particles  which correctly addresses the  noninertial features  of  rotating coordinates. We review it briefly and obtain the mass corrections to the chiral vector and axial-vector currents produced by this formulation.
\end{abstract}
	
	\maketitle
	
   \section{Introduction}
Quark-gluon plasma created in heavy-ion collisions form a novel phase of nuclear matter. The constituent quarks and gluons are deconfined, where the quarks can be considered as  massless, hence can be  right- or left-handed.  In the systems  composed of chiral fermions collective behavior generates the vector and axial-vector current terms  proportional to magnetic field known as the chiral magnetic effect \cite{kmw,fkw,kz} and the  chiral separation effect \cite{mz,jkr}. In currents, there also exist terms which are proportional to the vorticity. They lead to  the chiral vortical effect \cite{ss} and the local (spin) polarization effect \cite{lw,bpr,glpww}.  In these effects magnetic field and the vorticity of fluids play similar roles.	In heavy-ion collisions vortical effects generate the polarization of $\Lambda$ as measured in \cite{STAR}.   The dynamics of chiral fermions can be investigated by means of the covariant quantum kinetic equation in Minkowski space \cite{qBe,vge}. However, this transport equation lacks the similarity between  the magnetic field and the vorticity.  It depends explicitly on electromagnetic field strength but  the vorticity is  incorporated in it through  the fluid velocity \cite{glpww,hpy1,hsjlz} and the equilibrium distribution function \cite{css,hpy2}. Moreover, in the Wigner function approach  noninertial effects like the Coriolis force, do not  appear. To take into account  vorticity from the start, in  \cite{dk} we proposed and studied a modified quantum kinetic equation for  chiral  fermions.

To solve a transport equation   as the initial-value problem to guarantee that the distribution function is  acceptable, three-dimensional (3D)  kinetic equation is preferred \cite{bbgr,zh2,he}.  A methodical approach of introducing a 3D kinetic theory is  to integrate the four-dimensional (4D) one over the zeroth component of four-momentum, $p_0$ \cite{zh1,zh2}.  To perform this calculation one should choose a frame, but  not each possible choice yields an acceptable  3D kinetic theory. By integrating the 4D transport theory  proposed in \cite{dk} we accomplished  the 3D  kinetic theory  which is consistent with the chiral phenomena.  It is the unique consistent 3D semiclassical chiral kinetic equation which  does not depend explicitly on the  3D position vector $\bm x ,$ and possesses the Coriolis force. There exists a similar 3D formulation \cite{dkl}  which has been established from  the relativistic theory given in curved spacetime  \cite{lgmh}  by integrating  over $p_0,$ but  it depends on  $\bm x$ explicitly. A different approach of defining relativistic kinetic equations and the related 3D transport theories were discussed in \cite{cpww,gpw}.  A  consistent nonrelativistic kinetic formulation of chiral fermions both in the presence of electromagnetic fields and global rotations was given in  \cite{dky} by working directly in 3D.  Its starting point is the equivalence between fluid vorticity in hydrodynamical approach and angular velocity of  fluid  in the comoving frame \cite{sy}.  The kinetic theory of \cite{dky}  also depends  position explicitly, but it possesses the characteristic feature  of being acquired as the vanishing mass limit of the kinetic theory  of Dirac particles. Furthermore,  in this formulation similarity between the vorticity and magnetic field is explicit. Thus, it generates some new phenomena like the rotational analogue of Hall effect in  nonlinear transport of chiral plasma \cite{ofdek}.
	
Although in quark-gluon plasma the quarks are treated as massless, in reality they are massive.  Thus it is crucial to understand how nonvanishing fermion mass affects the   chiral transport phenomena.  The quantum kinetic equation of Wigner function was originally introduced  for   massive Dirac particles.  The Wigner function  can be decomposed in terms of the Clifford algebra  generators  whose coefficients are  scalar, pseudoscalar, vector, axial-vector and tensor fields. These fields satisfy a group of differential equations. In  the massless case  equations satisfied by the vector and axial-vector field components  decouple from the rest. However, for massive fermions one can choose different sets of field components to construct kinetic equations.  
One of the possible choices was given in \cite{gl}. Another choice was presented in  \cite{hhy}, where the vector and axial-vector fields were chosen as independent functions to construct kinetic equations whose massless limit can directly be acquired. An alternative  way of constructing transport equations from the quantum kinetic equation  was given in \cite{wsswr}. This approach has been extended to construct  kinetic theory of massive fermions in curved spacetimes \cite{lmh}.  In the  context of the semiclassical Wigner function method the study of  massive fermions was started in  \cite{F1,F2}.   By means of the  solutions of Dirac equation  they constructed Wigner function components at the zeroth-order in Planck constant.  Then,  inspired by the chiral one, they wrote the axial-vector field at the first-order in Planck constant and  the properties of polarization as well as pseudoscalar condensation for massive fermions  have been investigated. Mass corrections to chiral phenomena were also inspected by means of  equal-time kinetic equations  \cite{wgsz}. 
	
The aim of this work is to study massive spin-1/2 fermions in terms of two different semiclassical kinetic theory approaches whose common characteristics lie in the facts that the similarity between magnetic field and vorticity is manifest, and the noninertial features  like the Coriolis force are addressed correctly.  The first one is the 3D formulation presented in \cite{dky}.  We study the vector and axial-vector currents by paying attention to the terms which are linear in the magnetic field and the angular velocity.  We calculate  mass corrections to chiral effects. The second one is a  relativistic approach.  We propose to modify  the quantum kinetic equation of Dirac particles by means of   some enthalpy current dependent terms guided by the chiral formalism \cite{dk}.  But the modification of the quantum kinetic equation of Dirac particles differ from the massless  case
with  the terms which vanish in the massless limit. To derive  kinetic equations generated by the modified quantum kinetic equation  we mainly follow the method of \cite{wsswr}. After accomplishing the  kinetic equations of distribution functions  we integrate them over $p_0.$ Thereby  we establish the  3D transport equations in the helicity basis. This formulation leads to  the dispersion relation which is in accord with the one resulting from the Dirac Hamiltonian in rotating coordinates. Moreover,  the Coriolis force is generated.  We study mass corrections to chiral phenomena.
	
In Sec. \ref{sec3d},  the 3D formulation of \cite{dky} will   be reviewed briefly. The resulting  vector and axial-vector currents in the helicity basis  will be presented. Then the mass corrections to magnetic and vortical effects will be calculated  at zero temperature.  In Sec. \ref{secmqe}, we  discuss the modified quantum kinetic equation and obtain the transport equations. In comoving frame  we integrate them over $p_0$ after choosing the quantization direction appropriately. Thereby   the 3D kinetic equations are established. We discuss the particle number currents deduced from them  at zero temperature. In the last section we present discussions of our results and  open problems.

\section{3D Semiclassical Kinetic Theory }
\label{sec3d}
In \cite{dky}  dynamical features of Dirac particles  are studied by means of the semiclassical wave packets of free-particles, $e= 1$ and  free-antiparticles, $e=-1,$
\begin{equation}
\psi_{{\bm x}^\prime}^e (\bm{p}) = \sum_{\alpha=1,2} \xi_\alpha^e  u_\alpha^e  (\bm{p}, E) e^{-ei\bm{p}\cdot \bm{x}^\prime / \hbar},
\end{equation}
where $\bm{x}^\prime$ is the 3D position variable and $\bm{x},\bm{p}$ denote the  phase-space coordinates of wave packet center. \mbox{$E=\sqrt{\bm p^2 +m^2},$}  is the energy of  free particle of mass   $m.$  The spinors $u_\alpha^e ( \bm{p},E )$ denote the four linearly independent solutions of the Dirac equation and  $\xi_\alpha^e  $ are some constant coefficients. 

To expose the formulation of  \cite{dky}  briefly, let us suppress the index $e,$ and   deal  with   positive energy solutions. We define the one-form $\eta_0$ through 
 \begin{equation}
\int [dx^\prime]  \delta (\bm{x}-\bm{x}^\prime)\Psi^\dagger_{\bm{x}^\prime}\left( -i \hbar d -H_\ssD dt \right)\Psi_{\bm{x}^\prime}=\sum_{\alpha\beta}\xi^*_{\alpha} \eta_{0\alpha\beta} \xi_{\beta}.
\end{equation}
$\eta_0$ is a  $2\times 2$ matrix in ``spin indices.'' It  can be written as 
\begin{equation}
\eta_{0\alpha\beta}= \delta_{\alpha\beta}\bm{p}\cdot d\bm{x}  - \bm A_{\alpha\beta}\cdot d{\bm p} -H^\ssD_{\alpha\beta}dt  .
\label{et1}
\end{equation}
$H^\ssD_{\alpha\beta}$ is the projection of  Dirac Hamiltonian on the particle solutions, $u_\alpha \equiv u_\alpha^1,$ and 
\begin{equation}
\label{bgd}
\bm A_{\alpha\beta}= -i \hbar u^{\dagger}_\alpha (\bm p)\frac{\partial }{\partial {\bm p}} u_\beta (\bm p)=\hbar \frac{\bm{\sigma}_{\alpha\beta} \times \bm{p}}{2E(E+m)},
\end{equation}
is the matrix valued Berry gauge field where  $\bm \sigma$ are the Pauli spin matrices.  To establish  Hamiltonian formulation 
we define the following symplectic two-form by suppressing the matrix indices,
\begin{equation}
{\zeta}_{0t} = d\eta_0 \equiv {dp}_i \wedge{dx}_i - G -D_i H_D\ {dp}_i \wedge dt.
\end{equation}
The  repeated indices $i,j=1,2,3,$  are summed over. We introduced the covariant derivative:
\begin{equation}
\bm D \equiv \frac{\partial }{\partial \bm p}+\frac{i}{\hbar}[\bm A,\ ].
\end{equation}
Unit matrices are not explicitly written.
 $G=\frac{1}{2} {G_{ij}}{dp}_i \wedge {dp}_j$  is  the Berry curvature two-form which can be expressed in terms of the vector $\bm G$ as 
\begin{equation}
G_{ij} = \frac{\partial A_j}{\partial p_i}- \frac{\partial A_i}{\partial p_j}+\frac{i}{\hbar}[A_i,A_j]= {\epsilon}_{ijk}G_k
\label{eq:G}.
\end{equation}
By plugging (\ref{bgd}) into (\ref{eq:G}) one obtains
\begin{equation}
\bm G= \frac{\hbar m}{2E^3}\left( \bm{\sigma}+\frac{\bm{p}(\bm{\sigma}\cdot\bm{p})}{m(m+E)}\right)
\label{berrycurvature}.
\end{equation}
Up to now we dealt with the free Dirac particles. To discuss the electromagnetic  interactions and the rotation of  reference frame one needs to insert the appropriate gauge fields into the one-form. Once this is done the symplectic two-form matrix can be established as 
\begin{widetext}
\begin{eqnarray}
{\zeta}_t & = & {dp}_i \wedge {dx}_i + \frac{1}{2} \epsilon_{ijk} (q B_k + 2 {\cal E} \omega _k)\ {dx}_i \wedge {dx}_j -\frac{1}{2} \epsilon_{ijk} G_{k}\ {dp}_i \wedge {dp}_j \nonumber \\ 
&&+ \epsilon_{ijk} x_j\omega _k    \nu_m {dx}_i \wedge {dp}_m - \nu_i\ {dp}_i \wedge dt + \frac{1}{2}  \nu_i (\bm \omega  \times \bm x)^2 {dp}_i \wedge dt
\label{wtf}\\
&&+[q\bm E+ (\bm\omega \times \bm x)\times (q\bm B +{\cal E} \bm\omega )]_i\ {dx}_i\wedge dt .  \nonumber
\end{eqnarray}
\end{widetext}
The dispersion relation ${\cal E}$ and the related velocity $\bm v$ will be written explicitly in the sequel. 
$\bm E,\ \bm B$ are the electromagnetic fields and   $\bm \omega$ is 
the angular velocity of the rotating frame.
We consider nonrelativistic
global rotations, thus the linear velocity is bounded: $|\bm \omega \times \bm x|\ll c.$  

To establish the semiclassical transport equations consider
  the volume form
\begin{eqnarray}
\label{vftw}
{\Omega} &=& \frac{1}{3!} {\zeta}_t \wedge {\zeta}_t \wedge {\zeta}_t \wedge dt \nonumber\\
&=& 	\frac{1}{3!} {\zeta} \wedge {\zeta} \wedge {\zeta} \wedge dt .
\end{eqnarray}
The matrix valued ${\zeta}\equiv {\zeta}_t|_{dt=0}$ is the symplectic two-form  in the phase-space given by  $(\bm x,\bm p).$ 
One can equivalently write (\ref{vftw})   as 
\begin{equation}
\label{wfpf}
{\Omega}= {\zeta}_\ssh \ dV \wedge dt.
\end{equation}
$dV$ is the canonical  volume form and the phase space measure ${\zeta}_\ssh $  is the Pfaffian of the $(6\times 6)$ matrix,
\begin{equation}
\label{syma}
\begin{pmatrix}
\epsilon_{ijk} (q B_k + 2E \omega_k) & -\delta_{ij}+\nu_j(\bm x \times\bm \omega)_i \\
\delta_{ij}-\nu_i(\bm x \times\bm \omega)_j &\ -\epsilon_{ijk} G_{k}
\end{pmatrix}. 
\end{equation}

To attain the Liouville equation let us introduce  $i_{{v}},$ which is 
 the interior product of the vector field
\begin{equation}
\label{vf}
 v= \frac{\partial}{\partial t}+\dot{ {\bm x}}\frac{\partial}{\partial \bm{x}}+\dot{ {\bm p}}\frac{\partial}{\partial \bm{p}}.
\end{equation}
Note that $v$ and the time evolutions $(\dot{ {\bm x}},\dot{ {\bm p}})$ are  matrix-valued. As usual, we suppressed the unit matrix.  The Lie derivative 
$ {\cal{L}}_{ v}=i_{ v}  d + d i_{v} ,$ of the volume form 
yields the Liouville equation:
\begin{equation}
{\cal{L}}_{ v} {\Omega} =0.
\end{equation}
The Lie derivative can be calculated in two different ways.  By comparing  them one acquires the following  relation 
\begin{eqnarray}
\left( \frac{\partial {\zeta}_\ssh}{\partial t} + \frac{\partial}{\partial \bm x}\cdot  ( \dot{{ \bm x}}{\zeta}_\ssh) + \bm D\cdot ({\zeta}_\ssh \dot{ {\bm p}})\right) dV \wedge dt &=& \frac{1}{3!} d {{\zeta}_t}^3 .
\label{lievolume2}
\end{eqnarray}
Calculation of ${{\zeta}_t}^3$  yields   the explicit form of  Pfaffian and the time evolution  of phase space variables. Similar calculations can be carried out for antiparticles. Hence, by retrieving the particle/antiparticle index $e,$ we get 
\begin{eqnarray}
\zeta^e_\ssh &=&1+  \ \bm{G}^e  \cdot (e\bm{B} + 2 {\cal E}^e  \bm{\omega} ) - \bm \nu^e \cdot (\bm x \times \bm \omega )\nonumber\\
&& - (  \bm \nu^e   \cdot \bm G^e )[e \bm B \cdot (\bm x \times \bm \omega )],
\label{Pfaf}  \\
\dot{{ \bm x}}^e\zeta_\ssh^e&=&{\bm \nu}^{e}  [1 -\frac{1}{2} (\bm \omega  \times \bm x)^2]+  \ \bm E_\omega^e  \times \bm G^{e}  \nonumber\\
&&+  ({\bm \nu}^e  \cdot \bm G^e ) (e\bm B+ 2{\cal E}^e  \bm \omega )[1 -\frac{1}{2} (\bm \omega  \times \bm x)^2] \\
&&+ ( \bm \nu^e  \cdot \bm G^e ) [ (\bm x \times \bm \omega ) \times \bm E_\omega^e ]    ,\label{msxd} \nonumber \\
\zeta^e_\ssh \dot{ {\bm p}}^e&=& \bm E_\omega^e  + {\bm \nu}^e  \times (e\bm{B} + 2 {\cal E}^e  \bm{\omega })[1 -\frac{1}{2} (\bm \omega  \times \bm x)^2]  \label{mspd} \\
&& + \bm G^e  [\bm E_\omega^e  \cdot (e \bm{B} + 2{\cal E}^e  \bm{\omega })] - [ (\bm x \times \bm \omega ) \times \bm E_\omega^e  ] \times {\bm \nu}^e  . \nonumber
\end{eqnarray}
We introduced $\bm G^e=e\bm G,$ and 
 the effective electric fields
\begin{equation}
\label{efel}
\bm E_\omega^e  =  e\bm E+ (\bm\omega \times \bm x)\times (e\bm B +{\cal E}^e \bm\omega ). \nonumber
\end{equation}
We set the electric charge to unity, so that $q=e.$	The dispersion relation and the related velocity are obtained from the Hamiltonian of  charged Dirac particle under the influence of external electromagnetic fields, in a rotating frame as
\begin{equation}
\label{hwbc}
{\cal E}^e  =E [1-\bm G^e  \cdot (e\bm{B} + E\bm{\omega }) ].
\end{equation}
\begin{equation}
\bm \nu^e  =\frac{\bm p}{E} \left[ 1
+\bm G^e  \cdot \left(  2e \bm B+E\bm{\omega }\right) \right]
- \frac{ \hbar \chi e}{2E^3}   \ (e\bm B +E\bm \omega ) \bm{\sigma} \cdot \bm{p}  . \label{numas}
\end{equation}

Distribution functions are needed  to define particle number and current densities from one-particle quantities. Within this formalism distribution functions are matrices whose elements are  $f^e_{\alpha \beta}.$ In principle  all of the four elements can be nonvanishing. However,   we are mainly interested in mass corrections to the chiral kinetic theory. Hence, it is convenient to work in a basis which permits us to choose the distribution function to be diagonal:
\begin{equation}
\label{fdig}
 f^e=\begin{pmatrix}
f_\sspp^e & 0\\
0 & f_{\ssmm}^e
\end{pmatrix}.
\end{equation}
For massless fermions $f^e_\sschi;\ \chi= 1,-1,$ coincide with  distribution functions of the right- and left-handed particles/antiparticles. This basis can be called as the helicity basis:
Helicity matrix  is given by
\begin{equation}
\lambda =\frac{\bm{\Sigma}\cdot \bm{p}}{|\bm p|},
\end{equation}
where $\bm{\Sigma}=\begin{pmatrix} \bm{\sigma}&0\\0&\bm{\sigma}\end{pmatrix}$
is the spin matrix.  One can show that for the free particle/antiparticle  solutions 
\begin{equation}
u^{e\dagger}_\alpha \bm \Sigma u^e_\beta= e\bm \sigma_{\alpha \beta}.
\end{equation}
Recall that each $ u^e_\alpha$ is a 4-spinor. 
Now, let us change the basis by the $2 \times 2$ matrix $R$ which is defined to satisfy 
\begin{equation}
R^{\dagger} \bm \sigma R=
\begin{pmatrix}
1 & 0\\
0 & -1
\end{pmatrix}\bm n  ,
\end{equation}
where $\bm n$ is the spin quantization direction  in the rest frame of particle. 
Observe that by choosing
\begin{equation}
\label{hbas}
\bm n= \hat{\bm p},
\end{equation}
we can diagonalize the helicity matrix:
\begin{equation}
( u^e R)^\dagger (\frac{\bm{\Sigma}\cdot \bm{p}}{|\bm p|}) \ (u^e R)=
\begin{pmatrix}
1 & 0\\
 0 & -1
 \end{pmatrix} .
\end{equation}
Hence the choice (\ref{hbas}) is in accord with the  helicity basis. In this basis all the matrix valued quantities obtained above are diagonal. $\chi$ labels the diagonal elements.  For example 
 the Berry curvature  (\ref{berrycurvature}) is now diagonal whose elements are
\begin{equation}
\bm G^e_\chi  = \frac{\hbar e \chi \hat{\bm p}}{2E^2}  .
\end{equation}
 Thus, the dispersion  relation (\ref{hwbc}) turns out to be
\begin{equation}
\label{dish}
{\cal E}^\sschi_e  =E -  \frac{\hbar e \chi \hat{\bm p}}{2E} \cdot (e\bm{B} + E\bm{\omega }) .
\end{equation}
 The vorticity dependent shift in dispersion relation (\ref{dish}) is independent of mass. For chiral particles  the same term was obtained in    \cite{cetal} and  \cite{lgmh}.

We introduce the collisionless transport equations as 
	\begin{equation}
	\zeta^\sschi_e\frac{\partial f^e_\sschi}{\partial t} +  \dot{\bm x}^\sschi_e\zeta_e^\sschi\cdot \frac{\partial f^e_\sschi}{\partial \bm x} +  \zeta_e^\sschi\dot{\bm p}^\sschi_e \cdot \frac{\partial f^e_\sschi}{\partial \bm p}=0.
	\end{equation}
Now one can define particle number and current densities as
	\begin{eqnarray}
	n_\sschi(\bm x,t) &=& \int \frac{d^3p}{(2\pi \hbar)^3} \sum_e    \zeta^\sschi_e f^e_\sschi, \label{n,n}  \\
	\bm j_\sschi(\bm x,t) &=& \int \frac{d^3p}{(2\pi \hbar)^3} \sum_e ( \dot{\bm x}^\sschi_e\zeta_e^\sschi)  f^e_\sschi  .
	\label{n,j}
	\end{eqnarray}
	By setting  $\int \frac{d^3p}{(2\pi \hbar)^3}\sum_e \bm \partial_{p} \cdot(\zeta^\sschi_e \dot{{\bm p}}_e^\sschi f^e_\sschi )=0, $  the continuity equation follows,
	\begin{equation}
	\frac{\partial n_\sschi (\bm x,t)}{\partial t}  + \bm {\nabla} \cdot \bm j_e  (\bm x,t)   =0.
	\end{equation}
	The vector and axial vector currents are defined as
	\begin{equation}
	\bm j_\ssV (\bm x,t) = \sum_\sschi  \bm j_\sschi(\bm x,t),\ \ \ \ \bm j_\ssA (\bm x,t) = \sum_\sschi \chi \bm j_\sschi(\bm x,t).
	\end{equation}
They possess linear terms in $\bm B$ and $\bm \omega$ which can be expressed by
	\begin{equation}
	\bm j^{ B,  \omega}_{\ssA , \ssV} (\bm x,t) = \sigma_{\ssA , \ssV}^{\ssB} \bm B +\sigma_{\ssA , \ssV}^{ \omega} \bm \omega. 
	\end{equation}
	Let us choose  the equilibrium distribution function as the Fermi-Dirac distribution: 
	\begin{equation}
	\label{ffe}
	f^{\ssFD \sse}_{\scriptscriptstyle{({\cal E})}\sschi} =\frac{1}{e^{e[{\cal E}^\sschi_e  -\mu_\sschi]/T} +1}. 
	\end{equation}
After performing the   angular parts of the momentum space integrals we acquire
	\begin{eqnarray}
	\sigma_{\ssA , \ssV}^{\ssB}  &= & \frac{1}{6\pi^2 \hbar^2}\int d|\bm p|\left\{ \frac{|\bm p|^3}{E^3}f^{\ssFD}_{\ssA , \ssV}- 
	\frac{|\bm p|^3}{2E^2}\frac{\partial f^{\ssFD}_{\ssA , \ssV}}{\partial E}\right\} ,\label{sigB} \\
	\sigma_{\ssA , \ssV}^{\omega}   &= &\frac{1}{3 \pi^2 \hbar^2}\int d|\bm p|\left\{ \frac{|\bm p|^3}{E^2}f^{\ssFD}_{\ssA , \ssV}- 
	\frac{|\bm p|^3}{4E} \frac{\partial f^{\ssFD}_{\ssA , \ssV}}{\partial E}\right\} ,\label{sigO} 
	\end{eqnarray}
	where, after relabeling $\mu_\sspp \equiv \mu_\ssR$ and $\mu_\ssmm \equiv \mu_\ssL,$
	\begin{equation}
	\label{ffo}
	f^{\ssFD}_{\ssA , \ssV}=\sum_e\left\{\frac{1}{e^{e[ E -\mu_\ssR]/T} +1}\pm \frac{1}{e^{e[ E -\mu_\ssL]/T} +1}\right\}.
	\end{equation}
The upper and lower signs correspond to axial-vector and vector currents, respectively. These produce the  correct results for $m=0, $ as they have been calculated  explicitly in \cite{ofdek}.  For simplicity let us deal with $\mu_\ssR=\mu_\ssL=\mu.$ Then, the vector current coefficients vanish: $	\sigma_{  \ssV}^{\ssB,\omega}=0 .$     By performing the integrals in (\ref{sigB}) and (\ref{sigO}) at zero temperature one gets
\begin{eqnarray}
\sigma^{\ssB}_\ssA &=&\frac{1}{2\pi^2\hbar^2} \Big(\frac{\mu^2+3m^2}{\mu} - \frac{4 }{3} m \Big)\ \theta(\mu-m), \nonumber \\ 
\sigma_\ssA^{\omega}&=&\frac{1}{2\pi^2\hbar^2} \Big( \mu^2-m^2-\frac{4}{3} m^2 \ln{\left(\frac{\mu}{m}\right)}\Big)\ \theta(\mu-m).\nonumber
\end{eqnarray}
For small mass we may approximately set $\bm \nu \approx\hat{\bm p} $ and deal with $	f^{\ssFD}_{\ssA , \ssV} ,$ from the start.  At $T=0,$ this approximation  yields
	\begin{eqnarray}
	\sigma^{\ssB}_\ssA &=&\frac{1}{2\pi^2\hbar^2} \int_m^\mu dE \frac{\sqrt{E^2-m^2}}{E }
	\nonumber \\
	&\approx & \frac{1}{2\pi^2\hbar^2}\sqrt{\mu^2-m^2} \ \theta(\mu-m),
	\end{eqnarray}
	and
	\begin{eqnarray}
	\sigma_\ssA^{\omega} &=&\frac{1}{\pi^2\hbar^2} \int_m^\mu dE \sqrt{E^2-m^2} 
	\nonumber\\
	&\approx &\frac{\mu }{2\pi^2\hbar^2} \sqrt{\mu^2-m^2}\  \theta(\mu-m) .
	\end{eqnarray}
These are in accord with  the results acquired by field theoretic calculations using  Kubo formula for Dirac particles \cite{gmsw,bgb,lykubo}. Although for $T\neq 0$ it seems that small mass corrections differ,  a detailed study is needed. 

\section{Modified Quantum Kinetic Equation}
\label{secmqe}

Quantum kinetic equation of relativistic fluids has been derived for  charged  Dirac particles  in the presence of external electromagnetic fields \cite{qBe,vge}. It explicitly depends on the electromagnetic field strength $F_{\mu \nu}.$ But the vorticity tensor of  fluid which plays a role similar to electromagnetic field strength, shows up through the solution of kinetic equation.   However, (non)inertial features of relativistic vorticity can be inspected by means of the circulation tensor:
\begin{equation}
\label{circ}
W_{\mu \nu}= \partial_\mu W_\nu -\partial_\nu W_\mu ,
\end{equation}
where $W_\mu $ is  the enthalpy current and $\partial_\mu \equiv \partial / \partial x^\mu.$   The enthalpy current  is  defined  as $W_\mu =h u_\mu,$ where $u_\mu= d x_\mu /d\tau$ is the fluid four-velocity in the comoving frame and  $h$ is  the internal energy (enthalpy). 
Thus, the circulation tensor (\ref{circ})  turns out to be
\begin{equation}
W_{\mu \nu}= h \left(\partial_\mu u_\nu - \partial_\nu u_\mu \right)+(\partial_\mu h)u_\nu - (\partial_\nu h)u_\mu,
\end{equation}
In the rest frame of massive particles $h=m,$ so that in \cite{dk} we proposed  to  modify the quantum kinetic equation of massless particles by the substitution:
\begin{equation}
\label{subs}
F_{\mu \nu}\rightarrow F_{\mu \nu} +W_{\mu\nu}-h(\partial_{\mu}u_\nu -\partial_{\nu}u_\mu ).
\end{equation}
By virtue of the modified relativistic formulation  we established  a    3D semiclassical chiral kinetic theory which is consistent with anomalous chiral effects. Moreover, it possesses the Coriolis force and  does not depend explicitly on the  3D position vector $\bm x .$  
As we mentioned in the Introduction, it is the unique 3D theory which possesses all of  these properties. 

One can also  incorporate the noninertial effects in the quantum kinetic equation by considering it  in curved spacetime \cite{lgmh,lmh}. However, 
this is not in conflict with considering the modification (\ref{subs}) which is  in   Minkowski spacetime. In fact, in \cite{dkl} we showed that the modified chiral theory can be extended to curved spacetime.  However, depending on the   chosen frame and  observer, the modification terms may not give contribution to kinetic equation. We have shown that the formalism of \cite{lgmh} yields a consistent 3D chiral kinetic theory possessing Coriolis force but the vorticity terms depend explicitly  on  $\bm x .$ Thus,  studying noninertial effects in Minkowski spacetime by introducing modifications in terms of enthalpy  current has its own virtues. 

To incorporate the noninertial features of fluid vorticity into the Wigner function formalism of Dirac particles, we would like to modify their quantum kinetic equation  in a   similar manner. However, the modification  of quantum kinetic equation for massive particles differs from the chiral case with some terms which should survive for massive particles. We will show that  the  proposed modification generates correctly  the Coriolis force and  the dispersion relation of the Dirac particle coupled to electromagnetic fields in a rotating frame. 

We  present the modified quantum kinetic equation in a frame moving with the four-velocity $v_\mu;$ $v_\mu v^\mu =1,$  whose linear acceleration vanishes: $ v_\nu \partial^\nu v_\mu =0.$ We would like to modify the quantum kinetic equation with frame dependent terms adequate to consider the noninertial  features of relativistic vorticity. Thus we propose
\begin{equation}
\left[\gamma_\mu \left(\pi^\mu + \frac{i\hbar}{2} D^\mu \right)-m \right] W(x,p) = 0,
\label{qke}
\end{equation}
as the  quantum kinetic equation in a rotating frame of reference with
\begin{eqnarray}
D^{\mu} &\equiv & \partial^{\mu}-j_{0}(\Delta)\left[  F^{\mu\nu}  +w^{\mu\nu} \right]  \partial_{p \nu} , \label{Df}\\
\pi^{\mu} &\equiv &p^{\mu}-\frac{\hbar}{2} j_{1}(\Delta) \left[  F^{\mu\nu}  +w^{\mu\nu} \right]  \partial_{p \nu} ,\label{Pf}
\end{eqnarray}
where $\partial_p^\mu \equiv \partial / \partial p_\mu,$ and 
\begin{eqnarray}
w_{\mu\nu} =\frac{h}{2}( \partial_\mu v_\nu -\partial_\nu v_\mu )+(\partial_\mu h)v_\nu - (\partial_\nu h)v	_\mu.
\label{w_munu}
\end{eqnarray}
It is worth noting that  $w_{\mu\nu}$ is not the full circulation tensor (\ref{circ}). As we will show, our choice is dictated by the fact that  the correct dispersion relation and  Coriolis force should be generated  in 3D kinetic theory.

In this section the electric  charge $e$ is suppressed.
 \(j_{0}(x)\) and \(j_{1}(x)\)
are spherical Bessel functions in  \(\Delta \equiv \frac{\hbar}{2} \partial_{p} \cdot \partial_{x}\).  The space-time derivative \(\partial_{\mu}\) contained in \(\Delta\) acts on \(\left[  F^{\mu\nu}  +w^{\mu\nu} \right] ,\) but not on the Wigner function. In contrary  $ \partial_{p \nu}$ acts on the Wigner function, but not on  \(\left[  F^{\mu\nu}  +w^{\mu\nu} \right] .\)  

The decomposition of the Wigner function is written through the 16 generators of the Clifford algebra as
\begin{equation}
W=\frac{1}{4}\left(\mathcal{F}+i \gamma^{5} \mathcal{P}+\gamma^{\mu} \mathcal{V}_{\mu}+\gamma^{5} \gamma^{\mu} \mathcal{A}_{\mu}+\frac{1}{2} \sigma^{\mu \nu} \mathcal{S}_{\mu \nu}\right),
\label{wigner}
\end{equation}
where the coefficients $\mathcal{C}\equiv\left\{ \mathcal{F},\mathcal{P},\mathcal{V}_{\mu},\mathcal{A}_{\mu},\mathcal{S}_{\mu \nu}\right\} ,$  respectively, are the scalar, pseudoscalar, vector, axial-vector, and antisymmetric tensor components of the Wigner function.   
These fields can be expanded in powers of Planck constant:
\begin{equation}
\mathcal{C} =\sum_n\hbar^n\mathcal{C}^{(n)}.
\end{equation} 
We  deal with the semiclassical approximation where only the zeroth- and first-order fields are considered. Thus, to derive the equations which they satisfy, instead of  (\ref{Df}), (\ref{Pf}), we only need to deal with 
\begin{eqnarray}
\nabla^{\mu} &\equiv & \partial_{x}^{\mu}-\left[  F^{\mu\nu}  +w^{\mu\nu} \right]  \partial_{p \nu}  \\
\Pi^{\mu} &\equiv  &p^{\mu}-\frac{\hbar^2}{12}   \left[  \partial^\alpha F^{\mu\nu}  +\partial^\alpha w^{\mu\nu} \right] \partial_{p \alpha}\partial_{p \nu}. 
\end{eqnarray}
By  plugging   the decomposed Wigner function (\ref{wigner}) into the modified quantum kinetic equation (\ref{qke}),  one  derives the  equations satisfied by the  fields $\mathcal{C} , $ whose  real parts are
\begin{eqnarray}
\Pi\cdot \mathcal{V}-m \mathcal{F} =0,  \label{real1} \\
{\Pi_{\mu} \mathcal{F}-\frac{\hbar}{2} \nabla^{\nu} \mathcal{S}_{\nu \mu}-m \mathcal{V}_{\mu}=0},  \label{real2} \\
{-\frac{\hbar}{2} \nabla_{\mu} \mathcal{P}+\frac{1}{2} \epsilon_{\mu \nu \alpha \beta} \Pi^{\nu} S^{\alpha \beta}+m \mathcal{A}_{\mu}=0}, \label{real3} \\
{\frac{\hbar}{2} \nabla_{[\mu} \mathcal{V}_{\nu]}-\epsilon_{\mu \nu \alpha \beta} \Pi^{\alpha} \mathcal{A}^{\beta}-m \mathcal{S}_{\mu \nu}=0}, \label{real4} \\
\frac{\hbar}{2} \nabla \cdot \mathcal{A}+m \mathcal{P} =0, 
\label{real5}
\end{eqnarray}
and the imaginary parts are
\begin{eqnarray}
{\hbar \nabla \cdot \mathcal{V}=0}, \label{imag1} \\ 
{\Pi \cdot \mathcal{A}=0}, \label{imag2} \\ 
{\frac{\hbar}{2} \nabla_{\mu} \mathcal{F}+\Pi^{\nu} \mathcal{S}_{\nu \mu}=0},  \label{imag3} \\ 
{\Pi_{\mu} \mathcal{P}+\frac{\hbar}{4} \epsilon_{\mu \nu \alpha \beta} \nabla^{\nu} \mathcal{S}^{\alpha \beta}=0},  \label{imag4} \\ 
{\Pi_{[\mu} \mathcal{V}_{\nu]}+\frac{\hbar}{2} \epsilon_{\mu \nu \alpha \beta} \nabla^{\alpha} \mathcal{A}^{\beta}=0}.
\label{imag5}
\end{eqnarray}
Some of these equations can be employed to express a portion of the fields in terms of the others. Hence, depending on the choice of  independent set of fields one can follow different routes of studying the transport equations. Whatever the choices are, in the semiclassical approach one should first discuss  the general solutions at the zeroth-order in Planck constant and then at the higher orders.

\subsection{General solutions at ${\cal O}(\hbar )$    }
 
 Employing the solutions of  Dirac equation,  $u_\sschi^e ( \bm{p},E), $  the  Wigner function components at the  zeroth-order in Planck constant have been obtained in \cite{F1,F2,wsswr}  as follows
\begin{align}
\mathcal{F}^{(0)}=m \delta(p^2-m^2) \fV^{0} , \nonumber \\ 
\mathcal{V}_{\mu}^{(0)} = p_{\mu} \delta(p^2-m^2) \fV^{0},  \nonumber \\ 
\mathcal{A}_{\mu}^{(0)}=m s_{\mu} \delta(p^2-m^2) \fA^{0} , \label{zeroAS} \\
\mathcal{S}_{\mu \nu}^{(0)}= \Sigma_{\mu \nu}  \delta(p^2-m^2) \fA^{0} ,  \nonumber \\
{\mathcal{P}}^{(0)} =0, \nonumber 
\end{align}
where 
\begin{equation}
\Sigma_{\mu \nu}(x, p)=-\frac{1 }{m} \epsilon_{\mu \nu \alpha \beta} p^{\alpha} s^{\beta},
\end{equation}
is the dipole-moment tensor. 
Obviously, (\ref{zeroAS}) solve (\ref{real1})-(\ref{imag5}) by ignoring $\hbar $-dependent terms.    We choose the distribution functions in accord with (\ref{fdig}),  so that the scalar functions $\fV^0,\fA^0$ are given as
\begin{eqnarray}
\fV^{0}(x, p) &\equiv& 2 \sum_{e \sschi} \theta\left(e p^{0}\right) f_{\sschi}^{e}(x, p),  \label{fV}  \\
\fA^{0}(x, p) &\equiv& 2\sum_{e \sschi} s \theta\left(e p^{0}\right) f_{\sschi}^{e}(x, p). 
\label{fA}
\end{eqnarray}
$
s^\mu =\sum_e \theta (ep_0)s_e^\mu (\bm p, \bm n_e),
$
is the spin quantization direction four-vector satisfying  $s^2=-1,\  p \cdot s =0.$ It can be written in terms of the spin quantization direction in the rest frame of particle (antiparticle) $\bm n_e,$ as \cite{WGBook} 
\begin{equation}
s_e^\mu (\bm p, \bm n_e )= \left(\frac{\bm p \cdot \bm n_e}{m},   \ e\bm n_e + \frac{e\bm p \cdot \bm n_e }{m(E+m)}\bm p \right).
\label{s}
\end{equation}
Observe that the dipole-moment tensor
satisfies
$ \Sigma^{\mu \nu} \Sigma_{\mu \nu} =2,$ and the axial-vector component can be expressed as
\begin{equation}
\label{a0s}
\mathcal{A}_{\mu}^{(0)}=-\frac{1}{2} \epsilon_{\mu \nu \alpha \beta} p^{\nu} \Sigma^{\alpha \beta} \fA^{0}\delta(p^2-m^2)  .
\end{equation}
Multiplying (\ref{real1}), (\ref{real2}), respectively, with $m,$ $p_{\mu}$,  and keeping the \(\mathcal{O}\left(\hbar\right)\)  terms yield
\begin{equation}
(p^2-m^2) \mathcal{F}^{(1)}=\frac{1}{2} (F^{\mu \nu}+w^{\mu\nu}) \mathcal{S}_{\mu \nu}^{(0)}.
\end{equation}
Thus the scalar component of the Wigner function  at the first-order in $\hbar,$ can be written as
\begin{align}
\mathcal{F}^{(1)} = m \delta(p^2-m^2) \left(\fV^{1} + \frac{F^{\mu \nu}+w^{\mu\nu}}{2(p^2-m^2)} \mathcal{S}_{\mu \nu}^{(0)}\fA^{0}\right),
\label{firstF}
\end{align}
where $\fV^1$ is a scalar function.

Multiplying  (\ref{real3}),  (\ref{real4}) , respectively, with $\epsilon^{\mu \sigma \gamma \kappa} p_{\sigma},$ $m$ and joining them by employing   (\ref{imag2}),(\ref{imag3}), (\ref{zeroAS}), lead to
\begin{align}
(p^2-m^2) S^{\gamma \kappa (1)}=(F^{\gamma \kappa}+w^{\gamma \kappa}) \mathcal{F}^{(0)} . 
\end{align}
It is solved by 
\begin{align}
S^{\gamma \kappa (1)} = m \delta(p^2-m^2) \left( \Sigma^{1\gamma \kappa}  +  \frac{F^{\gamma \kappa}+w^{\gamma \kappa}}{(p^2-m^2)}\fV^{0}\right),
\label{firstS}
\end{align}
where $\Sigma^1_{\gamma \kappa}  $ can be thought of  as  the $\hbar$-order contribution to dipole-moment tensor. It satisfies the  constraint equation which follows from  (\ref{imag3}) by keeping terms at  \(\mathcal{O}\left(\hbar\right)\):
\begin{equation}
 \delta(p^2-m^2) p^\kappa\Sigma^1_{\gamma \kappa}=\frac{1}{2} \delta(p^2-m^2)\nabla_\gamma \fV^{0}.
\end{equation}

By substituting  the axial-vector with (\ref{a0s})  in (\ref{real5}) one finds
\begin{equation}
\mathcal{P}^{(1)}=\frac{1}{4 m} \epsilon^{\mu \nu \alpha \beta} \nabla_{\mu}\left[p_{\nu} \Sigma_{\alpha \beta} \fA^{0} \delta\left(p^{2}-m^{2}\right)\right]. 
\end{equation}

To acquire  the vector field $\mathcal{V}_{\mu}$ at    the  $\hbar$-order, we insert (\ref{zeroAS}) and (\ref{firstF})  into   (\ref{real2}): 
\begin{align}
\mathcal{V}_{\mu}^{(1)} &=\frac{p_{\mu}}{m} \mathcal{F}^{(1)}-\frac{\hbar}{2m} \nabla^{\nu} S_{\mu \nu}^{(0)} , \nonumber\\
&= \delta\left(p^{2}-m^{2}\right)\left[p_{\mu} \fV^{1}+\frac{1}{2} \nabla^{\nu}\left( \Sigma_{\mu \nu} \fA^{0}\right)\right],
\nonumber\\
&-\delta^{\prime}\left(p^{2}-m^{2}\right) \Big[\frac{1}{2} p_{\mu} (F^{\alpha \beta} +w^{\alpha \beta} )\Sigma_{\alpha \beta}\nonumber \\
&+\Sigma_{\mu \nu} \left(F^{\nu \alpha}+w^{\nu \alpha}\right) p_{\alpha}\Big] \fA^{0} .
\label{V1}
\end{align}

By plugging  (\ref{firstS}) into (\ref{real3}) one gets  the  axial vector field  at ${\mathcal{O}(\hbar)},$ as
\begin{align}
\mathcal{A}_{{\mu}}^{(1)} &=\frac{\hbar}{2 m} \nabla_{\mu} \mathcal{P}^{(0)}-\frac{1}{2 m} \epsilon_{\mu \nu \alpha \beta} p^{\nu} S^{{\alpha \beta (1)}},\nonumber\\ 
&=-\frac{1}{2} \epsilon_{\mu \nu \alpha \beta} p^{\nu} \Sigma^{1\alpha \beta}  \delta\left(p^{2}-m^{2}\right)\nonumber\\
&+\left(\tilde{F}_{\mu \nu}+\tilde{w}_{\mu \nu}\right) p^{\nu} \fV^{0} \delta^{\prime}\left(p^{2}-m^{2}\right),
\label{A1}
\end{align}
where  $\tilde{F}_{\mu \nu} =\frac{1}{2}\epsilon_{\mu \nu \alpha \beta} F^{\alpha \beta} $ and $ \tilde{w}_{\mu \nu}=\frac{1}{2}\epsilon_{\mu \nu \alpha \beta} w^{\alpha \beta},$ are the dual tensors. 

We solved a part of the equations (\ref{real1})-(\ref{imag5}) to write  the Wigner function components up to $\hbar$-order in terms of the undetermined distribution functions $\fV,\fA$ and spin quantization direction $\bm n.$  Now, by using the rest of  (\ref{real1})-(\ref{imag5}), we will establish the kinetic equations which they satisfy.

\subsection{Semiclassical kinetic quations }

By expanding (\ref{imag1}) and (\ref{imag5})  up to the first-order in $\hbar,$ one derives the  kinetic equations \cite{wsswr}:
\begin{align}
\delta\left(p^{2}-m^{2}\right) p \cdot \nabla \fV^{0}=0,
\label{eomV}   \\
\delta\left(p^{2}-m^{2}\right) p \cdot \nabla \fA^{0}=0,
\label{eomA} \\
\delta\left(p^{2}-m^{2}\right) \left(p \cdot \nabla \Sigma_{\mu \nu}-( F_{[\mu}^{\beta} + w_{[\mu}^{\beta} ) \Sigma_{\nu] \beta}\right)=0.
\label{eomSigma}
\end{align}
By inserting  (\ref{V1}) into (\ref{imag1}) and employing  the commutator 
\begin{equation}[\nabla_{\mu} , \nabla_{\nu}]= \{(\partial^\beta F_{\mu \nu}+\partial^\beta w_{\mu \nu}) - (F^{\beta \alpha} + w^{\beta \alpha}) (\partial_{p \alpha} \omega_{\mu \nu})  \} \partial_{p \beta}, 
\end{equation}
we obtain the following kinetic equation,   
\begin{widetext}
\begin{eqnarray}
\begin{aligned}
& \delta \left(p^{2}-m^{2}\right) \Big[p \cdot \nabla \fV
+\frac{\hbar}{4} \{(\partial^\beta F_{\mu \nu}+\partial^\beta w_{\mu \nu}) - (F^{\beta \alpha} + w^{\beta \alpha}) (\partial_{p \alpha} \omega_{\mu \nu})  \}  \Sigma^{\mu \nu} \partial_{p \beta} \fA^{0} &\\
&  +\frac{\hbar}{4} \{(\partial^\beta F_{\mu \nu}+\partial^\beta w_{\mu \nu}) - (F^{\beta \alpha} + w^{\beta \alpha}) (\partial_{p \alpha} \omega_{\mu \nu})  \} \fA^{0}\partial_{p \beta} \Sigma^{\mu \nu} \Big]& \\
&-\frac{\hbar}{2} \delta^{\prime}\left(p^{2}-m^{2}\right) \left(F^{\alpha \beta}+w^{\alpha \beta}\right)\Big[ \Sigma_{\alpha \beta} p \cdot \nabla \fA^{0} + \fA^{0} p \cdot \nabla \Sigma_{\alpha \beta}  \Big]=0,& 
\end{aligned} \label{ke0ms}
\end{eqnarray}
\end{widetext}
where  $\fV\equiv\fV^{0}+\hbar \fV^{1}.$   We can express the last term of (\ref{ke0ms}) in terms of the delta function, instead of its derivative: Let us  multiply the zeroth-order equation  (\ref{eomSigma}) by $\left(F^{\alpha \beta}+w^{\alpha \beta}\right),$ use $\left(F^{\mu \nu}+w^{\mu \nu}\right) (F_{[\mu}^{\beta} + w_{[\mu}^{\beta} ) \Sigma_{\nu] \beta}=0,$ and then take its derivative with respect to momentum. Contracting the resultant equation   with momentum  results in
\begin{widetext}
\begin{equation}
-\delta^{\prime}\left(p^{2}-m^{2}\right) \left(F^{\alpha \beta}+w^{\alpha \beta}\right) p \cdot \nabla \Sigma_{\alpha \beta} =\frac{\delta\left(p^{2}-m^{2}\right)}{2 p^2}\Big\{  \left(F^{\alpha \beta}+w^{\alpha \beta}\right)p \cdot \nabla \Sigma_{\alpha \beta}
+p^\mu p^\nu \partial_{p \nu} \left[\left(F^{\alpha \beta}+w^{\alpha \beta}\right)\nabla_\mu\Sigma_{\alpha \beta}\right]
\Big\}. 
\end{equation}
Thus the kinetic equation (\ref{ke0ms})  can equivalently be written as
\begin{eqnarray}
\begin{aligned}
& \delta \left(p^{2}-m^{2}\right) \Big\{p \cdot \nabla \fV
+\frac{\hbar}{4} \Big[(\partial^\beta F_{\mu \nu}+\partial^\beta w_{\mu \nu}) - (F^{\beta \alpha} + w^{\beta \alpha}) (\partial_{p \alpha} \omega_{\mu \nu})  \Big]  \Sigma^{\mu \nu} \partial_{p \beta} \fA^{0} &\\
&  +\frac{\hbar}{4} \{(\partial^\beta F_{\mu \nu}+\partial^\beta w_{\mu \nu}) - (F^{\beta \alpha} + w^{\beta \alpha}) (\partial_{p \alpha} \omega_{\mu \nu})  \} \fA^0\partial_{p \beta} \Sigma^{\mu \nu} & \\
&  +\frac{\hbar}{2p^2}\fA^0\left(F^{\alpha \beta}+w^{\alpha \beta}\right)p \cdot \nabla \Sigma_{\alpha \beta}
+\frac{\hbar p^\mu p^\nu }{2p^2}\fA^0\partial_{p \nu} \left[\left(F^{\alpha \beta}+w^{\alpha \beta}\right)\nabla_\mu\Sigma_{\alpha \beta}\right]
\Big\}&\\
&-\frac{\hbar}{2} \delta^{\prime}\left(p^{2}-m^{2}\right) \left(F^{\alpha \beta}+w^{\alpha \beta}\right)\Sigma_{\alpha \beta} p \cdot \nabla \fA^{0} =0.& 
\end{aligned} \label{ke1ms}
\end{eqnarray}
\end{widetext}
There are four unknown  functions in the general solutions of Wigner function components. Hence, we need to obtain some other  kinetic equations: At the $\hbar^2$-order (\ref{imag5}) yields
\begin{equation}
p_{[\mu} {\mathcal{V}_{{\nu]}}^{(2)}}+{\Pi^{(2)}_{[\mu}} {\mathcal{V}_{{\nu]}}^{(0)}}+\frac{\hbar}{2} \epsilon_{\mu \nu \alpha \beta} \nabla^{\alpha} \mathcal{A}^{\beta (1)}=0.
\label{secondeq}
\end{equation}
$\mathcal{V}^{(2)}_{{\nu}}$ can be read from (\ref{real2}) and $\mathcal{A}_{\mu}^{(1)}$  is given by (\ref{A1}). Inserting them into  (\ref{secondeq}) and  using (\ref{eomA}), (\ref{eomSigma}), one finds
\begin{widetext}
\begin{eqnarray}
\begin{aligned}
& \delta\left(p^{2}-m^{2}\right)\left[p \cdot \nabla \left(\Sigma_{\mu \nu}\fA^{0} +\hbar \Sigma^1_{\mu \nu}\right)-\left( w_{[\mu}^{\alpha}+F_{[\mu}^{\alpha} \right)\left(\Sigma_{\nu ] \alpha}\fA^0
+\hbar \Sigma_{\nu ] \alpha}^1\right)
+\frac{\hbar}{2}\left(\partial_{ \alpha} F_{\mu \nu}+\partial_{ \alpha} w_{\mu \nu}\right) \partial_{p}^{\alpha} \fV\right] \\
&-\hbar \delta^{\prime}\left(p^{2}-m^{2}\right) \left(F_{\mu \nu}+w_{\mu \nu}\right) p \cdot \nabla \fV +\hbar \delta^{\prime}\left(p^{2}-m^{2}\right) p^\alpha \left(F_{\alpha \rho}+w_{\alpha \rho}\right) \left(\partial_p^\rho w_{\mu \nu} \right)\fV =0.
\end{aligned} \label{ke2}
\end{eqnarray}
By contracting it  with $\frac{1}{2}\epsilon^{\mu \nu \alpha \beta} p_{\alpha},$  (\ref{ke2})  can also be expressed as 
\begin{eqnarray}
\begin{aligned}
& \frac{\delta\left(p^{2}-m^{2}\right)}{m}\left[ p \cdot \nabla (s^{\beta }\fA^0)- \left( w^{ \beta \alpha}+F^{\beta \alpha} \right) s_{\alpha} \fA^0\right] 
+\hbar \frac{ \delta\left(p^{2}-m^{2}\right) }{2 p^2}\epsilon^{\mu \nu \alpha \beta} p_{\alpha}\left[ p \cdot \nabla \Sigma^1_{\mu \nu} -\left( w_{[\mu}^{\alpha}+F_{[\mu}^{\alpha} \right) \Sigma_{\nu ] \alpha}^1 \right]
\\& +\hbar \frac{\delta\left(p^{2}-m^{2}\right)}{2 p^2} p_{\alpha} \left(\partial_{ \sigma} \tilde{F}^{\alpha \beta}+\partial_{ \sigma} \tilde{w}^{\alpha \beta}\right) \partial_{p}^{\sigma} \fV  \\
&-\hbar\frac{\delta^{\prime}\left(p^{2}-m^{2}\right)}{p^{2}} p_{\alpha} \left(\tilde{F}^{\alpha \beta}+\tilde{w}^{\alpha \beta}\right) p \cdot \nabla \fV 
+\hbar \frac{\delta^{\prime}\left(p^{2}-m^{2}\right)}{p^{2}}  p_\alpha p^{\sigma} \left(F_{\sigma \gamma}+w_{\sigma \gamma}\right) \left(\partial_p^\gamma \tilde{w}^{\alpha \beta} \right)\fV =0.& 
\end{aligned} 
\end{eqnarray} 
Let us focus on its projection in the spin  quantization direction  by multiplying it with $(-s^\beta/m),$ which is equivalent to taking contraction of  (\ref{ke2})  with $\frac{1}{2} \Sigma^{\mu \nu} :$ 
\begin{eqnarray}
\begin{aligned}
& \delta\left(p^{2}-m^{2}\right)\left[p \cdot \nabla \fA+\frac{\hbar}{4}\Sigma^{\mu \nu} \left(\partial_{ \alpha} F_{\mu \nu}+\partial_{ \alpha} w_{\mu \nu}\right) \partial_{p}^{\alpha} \fV \right]\\
&-\frac{\hbar}{2} \delta^{\prime}\left(p^{2}-m^{2}\right) \Sigma^{\mu \nu} \left(F_{\mu \nu}+w_{\mu \nu}\right) p \cdot \nabla \fV \\
&+\frac{\hbar}{2} \delta^{\prime}\left(p^{2}-m^{2}\right)  \Sigma^{\mu \nu} p^\alpha \left(F_{\alpha \rho}+w_{\alpha \rho}\right) \left(\partial_p^\rho w_{\mu \nu} \right)\fV =0,&
\end{aligned} \label{ke20ms}
\end{eqnarray}
where we introduced
\begin{equation}
\fA\equiv \fA^0+ \frac{\hbar}{2} \Sigma^{\mu \nu}\Sigma^1_{\mu \nu}.
\end{equation}
By summing and subtracting (\ref{ke1ms}) and (\ref{ke20ms}) we get
\begin{eqnarray}
\begin{aligned}
& \delta \left(p^{2}-m^{2} \mp \frac{\hbar}{2}  \Sigma_{\mu \nu} \left(F^{\mu \nu}+w^{\mu \nu}\right) 
\right) \Big\{p \cdot \nabla \left(\fV \pm \fA\right) &  \\
& \pm  \frac{\hbar}{4}\Sigma^{\mu \nu} \left(\partial_{x \alpha} F_{\mu \nu}+\partial_{x \alpha} w_{\mu \nu}\right) \partial_{p}^{\alpha}  \left(\fV \pm \fA\right) 
- \frac{\hbar}{4} \Sigma^{\mu \nu}  ( F^{\beta \alpha }+w^{\beta \alpha}) (\partial_{p \alpha} w_{\mu \nu}) \partial_{p\beta}     \fA \Big\}  +{\cal C}_1\pm{\cal C}_2 =0. &
\end{aligned} \label{kms}
\end{eqnarray}
${\cal C}_1, {\cal C}_2$ designate  the terms which do not contain derivatives of $\fA$ or  $\fV :$ 
\begin{eqnarray}
{\cal C}_1&=&   \delta \left(p^{2}-m^{2}\right) \frac{\hbar \fA}{4} \Big\{
\{(\partial^\beta F_{\mu \nu}+\partial^\beta w_{\mu \nu}) - (F^{\beta \alpha} + w^{\beta \alpha}) (\partial_{p \alpha} \omega_{\mu \nu})  \} \partial_{p \beta} \Sigma^{\mu \nu} \nonumber \\  &&
+\frac{2}{p^2}\left(F^{\alpha \beta}+w^{\alpha \beta}\right)p \cdot \nabla \Sigma_{\alpha \beta} 
+\frac{2p^\mu p^\nu }{p^2}\partial_{p \nu} \left[\left(F^{\alpha \beta}+w^{\alpha \beta}\right)\nabla_\mu \Sigma_{\alpha \beta} \right] \Big\}, \\
{\cal C}_2&=&\frac{\hbar}{2} \delta^{\prime}\left(p^{2}-m^{2}\right)  \Sigma^{\mu \nu} p^\alpha \left(F_{\alpha \rho}+w_{\alpha \rho}\right) \left(\partial_p^\rho w_{\mu \nu} \right)\fV . \nonumber
\end{eqnarray}
\end{widetext}
The zeroth-order particle distribution functions given in (\ref{fV}), (\ref{fA}), can be extended as follows,
\begin{eqnarray}
\frac{1}{2}(\fV+\fA)= 2\sum_e \theta\left(ep^{0}\right) f_\ssR^{e}(x, p) , \\ 
\frac{1}{2}(\fV-\fA)=2 \sum_e \theta\left(ep^{0}\right) f_\ssL^{e}(x, p),
\end{eqnarray}
where we relabeled $f_\sspp \equiv f_\ssR$ and $f_\ssmm \equiv f_\ssL.$

An alternative form of kinetic equations is given in Appendix \ref{apalt}.

\section{3D kinetic equations in the comoving frame}

In Sec. \ref{secmqe}, we modified the  quantum kinetic equation in terms of  $w_{\mu\nu}$ given in (\ref{w_munu}), where $v_\mu$ is an  arbitrary frame velocity.
Let us now deal with the comoving frame by setting $v_\mu=u_\mu.$ As we have already mentioned, we consider vanishing  linear acceleration, $ u_\nu \partial^\nu u_\mu =0.$  
Let us choose the distribution function $ f_{\sschi}$ as
\begin{equation}
f^{\scriptscriptstyle{FD}}_{\scriptscriptstyle{(u\cdot p)}\sschi} =\sum_{e}    \frac{2\theta(ep^0)}{ e^{e( u \cdot p - \mu_\sschi) / T } +1  } \cdot
\label{f_fd} 
\end{equation}
Now one can observe that to satisfy the zeroth-order equations (\ref{eomV}),  (\ref{eomA}), the fluid velocity should fulfill the condition $ \partial_\nu u_\mu =-\partial_\mu u_\nu.$ Therefore we can express  the vorticity tensor as 
\begin{equation}
\partial_\mu u_\nu =\epsilon_{\mu \nu \rho \sigma}u^\rho \omega^{\sigma },
\end{equation}
where  $\omega_{\mu }$ is  the vorticity of fluid. The  
internal energy is $h=u\cdot p.$ Thereby (\ref{w_munu}) can be written as follows
\begin{eqnarray}
w_{\mu\nu}^{\scriptscriptstyle{(CF)}} &=&u\cdot p  \ \epsilon_{\mu \nu \sigma \rho} u^\sigma \omega^{\rho} + p^{\alpha}  (u_\nu \ \epsilon_{\mu \alpha \sigma \rho} - u_\mu \ \epsilon_{\nu \alpha \sigma \rho} ) u^\sigma \omega^{\rho} 
\nonumber \\
&=&\epsilon_{\mu \nu \sigma \rho} \ \omega^{\rho}  \ \{ 2 (u \cdot p)  \ u^\sigma -  p^\sigma \}   ,
\end{eqnarray} 
where we employed the Schouten identity:
$
u_\nu \epsilon_{\mu \alpha \sigma \rho}= - (u_\mu \epsilon_{ \alpha\nu \sigma \rho} + u_\alpha \epsilon_{ \nu \mu \sigma \rho} + u_\sigma \epsilon_{ \rho \nu \mu \alpha} + u_\rho \epsilon_{ \nu \mu \alpha \sigma} ).
$

In order to obtain a 3D transport equation by integrating (\ref{kms}) over $p_0,$ we need to specify $s^{\mu}$ in accord with the equation (\ref{eomSigma}).  Because of expressing it  as in (\ref{s}), this is equivalent to solve (\ref{eomSigma}) for  $\bm n_e.$  Although the solution will be  in the form  $\bm n_e= \bm n_e ( p,F,w),$  to have an idea about the 3D transport equations let us ignore its   $F$ as well as $w$ dependence and choose it adequate to the helicity basis: $\bm n_e =  \hat{\bm p}.$ Obviously, 3D spin quantization direction for massive fermions need not to be in the direction of  momentum. However, as we have seen in Sec. \ref{sec3d}, helicity basis is useful to obtain the mass corrections to the chiral effects.  Therefore, the spin quantization direction four-vector becomes
\begin{equation}
\label{smu}
s^{\mu}_{e} (\bm p ,\bm p) =  \left(\frac{|\bm p|}{m},  e\frac{E}{m}\hat{\bm p}\right).
\end{equation}
Because of this choice when we integrate the kinetic equation (\ref{kms}) over $p_0,$
we need to ignore the terms which are quadratic in fields. To perform the integral one should first solve the mass-shell condition for $p_0:$
\begin{equation}\label{delp}
p_0^2-E^2+\chi \hbar\frac{ \epsilon^{\mu \nu \alpha \beta} p_\alpha s_\beta}{2 m} (F_{\mu \nu}+w^{\scriptscriptstyle{(CM)}}_{\mu \nu})=0.\end{equation}
We work in the frame $u_\mu=(1,\bm 0),\  \omega^\mu= (0, \bm \omega),$  hence  by substituting $s_\mu$ with
 (\ref{smu}),  one can easily see that (\ref{delp})  dictates the  dispersion relation 
\begin{eqnarray}
\label{hwbc2}
p_0=e E-\frac{\hbar \chi}{2E} \hat{\bm p} \cdot  ( \bm B +E \bm \omega )\equiv e 
{\cal E}^\sschi_e .
\end{eqnarray}
This is in accord with (\ref{dish}). Thus, we establish the 3D transport equation
\begin{equation}
\big( \sqrt{\eta}_{\, \sschi}^{\, e} \frac{\partial }{\partial t} + (\sqrt{\eta} \dot{{\bm x}})_{\sschi}^{e}  \cdot \frac{\partial }{\partial \bm{x}} + (\sqrt{\eta}  \dot{\bm p})_{\sschi}^{e} \cdot\frac{\partial }{\partial \bm{p}}\big) f_{\sschi}^{e} (t,\bm x,\bm p)=0,
\label{boltzman} 
\end{equation}
where
\begin{eqnarray}
\sqrt{\eta}_{\, \sschi}^{\, e}    &=&  1- \frac{\hbar \chi }{2m^2} \Big(\hat{\bm p} \cdot  \bm B - \frac{m^2+E^2}{E} e\hat{\bm p} \cdot \bm \omega \Big) \label{3e1}, \nonumber\\ 
(\sqrt{\eta} \dot{{\bm x}})_{\sschi}^{e}  &=& \frac{\bm p}{E} -\hbar \chi\hat{\bm p}\Big(   \frac{\bm p \cdot \bm B }{2Em^2}- \frac{\bm p \cdot  \bm B }{2E^3} + e \frac{\bm p\cdot \bm \omega}{2m^2}  \Big),\label{3e2} \\
(\sqrt{\eta}  \dot{\bm p})_{\sschi}^{e}&=& e \bm E+ 2 \bm p \times \bm \omega
+ \frac{\bm p \times e \bm B}{E}. \nonumber
\end{eqnarray}
To acquire the vector and axial-vector currents,  we  insert  (\ref{3e2}) into the definition (\ref{n,j}). Now the terms which are parallel to $\bm B$ and $\bm \omega$ can be shown to be 
\begin{equation}
\bm j^{ B,  \omega}_{\ssA , \ssV} (\bm x,t) = \bar{\sigma}_{\ssA , \ssV}^{\ssB} \bm B +\bar{\sigma}_{\ssA , \ssV}^{ \omega} \bm \omega,
\end{equation}
with
\begin{eqnarray}
\bar{\sigma}_{\ssA , \ssV}^{\ssB}  &= & \frac{1}{6\pi^2 \hbar^2}\int d|\bm p|\left\{ -\frac{|\bm p|^5}{m^2 E^3}f^{\ssFD}_{\ssA , \ssV} 
-\frac{|\bm p|^3}{E^2}\frac{\partial f^{\ssFD}_{\ssA , \ssV}}{\partial E}\right\} , \\
\bar{\sigma}_{\ssA , \ssV}^{\omega}   &= &\frac{1}{6 \pi^2 \hbar^2}\int d|\bm p|\left\{ -\frac{|\bm p|^3}{m^2}f^{\ssFD}_{\ssA , \ssV}- 
\frac{|\bm p|^3}{E} \frac{\partial f^{\ssFD}_{\ssA , \ssV}}{\partial E}\right\}.
\end{eqnarray}
 $ f^{\ssFD}_{\ssA , \ssV}$ are  defined in (\ref{ffo}).  For $\mu_\ssR=\mu_\ssL=\mu,$ at $T=0,$ the vector current coefficients vanish and  the axial-vector current coefficients can be calculated as
\begin{eqnarray}
\bar{\sigma}_{\ssA}^{\ssB} &=& \frac{1}{2 \pi^2 \hbar^2} \Big(\mu-\frac{\mu^3+8m^3}{9 m^2}\Big) \ \theta(\mu-m), \\
\bar{\sigma}_{\ssA}^{\omega} &=& \frac{1}{4 \pi^2 \hbar^2} \Big(\mu^2-\frac{\mu^4}{6m^2} - \frac{5 m^2}{2}\Big) \ \theta(\mu-m). 
\end{eqnarray}
These reflect the fact that in this formulation the zero mass limit cannot be achieved  directly. In  relativistic Wigner function formalism currents are  calculated  by integrating  the equilibrium vector and axial-vector fields  over the 4-momentum space. For massless fermions  the chiral effects should not depend on the kinetic theory formalism. However, in the massive case equilibrium distributions  depend on the formalism adopted to define the kinetic equations \cite{gl,hhy,wsswr,lmh}.  Hence there is no consensus about the mass corrections to chiral effects. Usually one discusses the small mass limit as we have done in Sec. \ref{sec3d}. However, the formalism which we adopted suits well with large  mass \cite{hhy}. Moreover, we have chosen the helicity basis (\ref{smu}) obtained by oversimplifying the  kinetic equation  (\ref{eomSigma}). Nevertheless, this choice permitted us to show that our formalism yields the 3-dimensional  kinetic theory with the correct dispersion relations (\ref{hwbc2}) and the Coriolis force. 

\section{Discussions }
We studied the semiclassical kinetic theories of Dirac particles  within two different approaches. First, the 3D formalism of Dirac particles  \cite{dky}, which manifestly exhibits the magnetic field, vorticity similarity is studied in helicity basis. We derived the axial-vector and vector currents for massive spin-1/2 particles. One of the distinguishing properties of this 3D formalism  is the fact that massless limit can be reached effortlessly.  Actually, one can easily observe that the currents (\ref{sigB}),(\ref{sigO}),  generate the chiral magnetic and vortical effects correctly.  We calculated   the  axial-vector current at zero temperature. In the small mass limit  it is consistent with the Kubo formalism based calculations \cite{lykubo}. 

Dirac fermions in external electromagnetic fields can be considered as relativistic fluids  described by the Wigner function satisfying the quantum kinetic equation which involves the electromagnetic field strength. For being able to take into account  the noninertial characteristics of vorticity  we modified this kinetic equation with  terms depending on the four-velocity of  rotating  frame.  We studied the  equations 
satisfied by  the  Wigner function components and  established  the semiclassical relativistic kinetic equations of  the scalar functions and  spin degrees of freedom. Kinetic equations of the scalar functions are integrated  over $p_0$ in a comoving frame by choosing the spin direction adequate  to  helicity basis. It is  shown that the Coriolis force and the  correct energy dispersion relation are generated. In addition,  the coefficients of  magnetic and vortical terms  in axial-vector and vector currents are calculated at zero temperature. Unfortunately, the massless limit cannot be acquired directly within the approach which we have adopted to establish the semiclassical  kinetic equations. Obviously, the resulting 3D model relies on the choice of spin quantization direction given in (\ref{smu}).    For  having a better understanding of  3D kinetic equations resulting from  the relativistic ones, one should study the solution of (\ref{eomSigma}) without ignoring the external electromagnetic fields and vorticity.

The original quantum kinetic equation was derived from the Dirac equation of charged particles coupled to the electromagnetic  vector field $A_\mu (x).$  Hence, to consider the noninertial features of vorticity  one can think to add a term proportional to the four-velocity $u_\mu (x) $ into the Dirac equation. It can be added only with a coefficient possessing the dimension of mass. Although internal  energy, $h,$ has mass dimension, in general it depends on momentum $p_\mu.$ The unique possibility is to choose the coefficient equal to $m.$  However, this choice does not yield an enthalpy current which we need to take into account the noninertial  properties correctly.  Hence,  the modification which we propose  does not seem to be generated by some gauge fieldlike terms coupled to Dirac particles.

\begin{acknowledgments}
	This work is supported by the Scientific and Technological Research Council of Turkey (T\"{U}B\.{I}TAK) Grant No. 117F328.
\end{acknowledgments}

\appendix

\renewcommand{\theequation}{\thesection.\arabic{equation}}
\setcounter{equation}{0}

\section{Alternative form of kinetic equations}
\label{apalt}

 All physical quantities are  defined through integrals over momentum variables. Therefore, they are defined up to  partial integrations. 
 Consequently, we may get rid of the derivative of delta function in (\ref{ke20ms}). First express it as
$
\frac{1}{2} \partial_p^\alpha \delta \left(p^{2}-m^{2}\right)= p^\alpha \delta^{\prime}\left(p^{2}-m^{2}\right), 
$
then perform the partial integration in the last term of  (\ref{ke20ms}), which results in 
\begin{widetext}
\begin{eqnarray}
\begin{aligned}
-\frac{\hbar}{4} \delta\left(p^{2}-m^{2}\right) \left(F_{\alpha \beta}+w_{\alpha \beta}\right)  \left(\partial_p^\beta w_{\mu \nu} \right)  \Big[  \Sigma^{\mu \nu} \partial_p^\alpha \fV 
+( \partial^\alpha_p \Sigma^{ \mu \nu})\fV\Big]. \nonumber
\end{aligned}
\end{eqnarray}
Thus  (\ref{ke20ms}) can equivalently  be written as
\begin{eqnarray}
\begin{aligned}
& \delta\left(p^{2}-m^{2}\right)\Big\{ p \cdot \nabla \fA+\frac{\hbar}{4}\Sigma^{\mu \nu} \left(\partial_{ \alpha} F_{\mu \nu}+\partial_{ \alpha} w_{\mu \nu}\right) \partial_{p}^{\alpha} \fV 
-\frac{\hbar}{4} \Big[  \Sigma^{\mu \nu} \left(F_{\alpha \beta}+w_{\alpha \beta}\right) \left(\partial_p^\beta w_{\mu \nu} \right)\partial_p^\alpha \fV &\\
& +
\left(F_{\alpha \beta}+w_{\alpha \beta}\right)( \partial^\alpha_p \Sigma^{\mu \nu}) \left(\partial_p^\beta w_{\mu \nu} \right)\fV\Big]\Big\} -\frac{\hbar}{2} \delta^{\prime}\left(p^{2}-m^{2}\right) \Sigma^{\mu \nu} \left(F_{\mu \nu}+w_{\mu \nu}\right) p \cdot \nabla \fV =0. &
\end{aligned} \label{ke2mss}
\end{eqnarray}
By summing and subtracting (\ref{ke1ms}) and (\ref{ke2mss}) we get
\begin{eqnarray}
\begin{aligned}
& \delta \left(p^{2}-m^{2} \mp \frac{\hbar}{2}  \Sigma_{\mu \nu} \left(F^{\mu \nu}+w^{\mu \nu}\right) 
\right) \Big\{p \cdot \nabla \left(\fV \pm \fA\right) &  \\
& \pm  \frac{\hbar}{4}\Sigma^{\mu \nu} \left(\partial_{x \alpha} F_{\mu \nu}+\partial_{x \alpha} w_{\mu \nu}\right) \partial_{p}^{\alpha}  \left(\fV \pm \fA\right) 
\mp \frac{\hbar}{4} \Sigma^{\mu \nu}  ( F^{\beta \alpha }+w^{\beta \alpha}) (\partial_{p \alpha} w_{\mu \nu}) \partial_{p\beta}    \left(\fV \pm \fA\right) \Big\}  +\tilde{\cal C}_1\pm \tilde{\cal C}_2 =0. &
\end{aligned} \label{kmss} 
\end{eqnarray}
$\tilde{\cal C}_1, \tilde{\cal C}_2$ indicate the terms which do not contain derivatives of $\fA$ or  $\fV :$ 
\begin{eqnarray}
\tilde{\cal C}_1&=&   \delta \left(p^{2}-m^{2}\right) \frac{\hbar \fA}{4} \Big\{
\{(\partial^\beta F_{\mu \nu}+\partial^\beta w_{\mu \nu}) - (F^{\beta \alpha} + w^{\beta \alpha}) (\partial_{p \alpha} \omega_{\mu \nu})  \} \partial_{p \beta} \Sigma^{\mu \nu} \nonumber \\  &&
+\frac{2}{p^2}\left(F^{\alpha \beta}+w^{\alpha \beta}\right)p \cdot \nabla \Sigma_{\alpha \beta} 
+\frac{2p^\mu p^\nu }{p^2}\partial_{p \nu} \left[\left(F^{\alpha \beta}+w^{\alpha \beta}\right)\nabla_\mu \Sigma_{\alpha \beta} \right] \Big\}, \\
\tilde{\cal C}_2&=&-\frac{\hbar}{4}\delta \left(p^{2}-m^{2}\right)
\left(F_{\alpha \beta}+w_{\alpha \beta}\right)( \partial^\alpha_p \Sigma^{\mu \nu}) \left(\partial_p^\beta w_{\mu \nu} \right)\fV. \nonumber
\end{eqnarray}
\end{widetext}
These kinetic equations do not involve derivatives of delta functions.

\newcommand{\PRL}{Phys. Rev. Lett. }
\newcommand{\PRB}{Phys. Rev. B }
\newcommand{\PRD}{Phys. Rev. D }

\end{document}